\documentclass[aps,prd,twocolumn,showpacs,eqsecnum,nofootinbib,amsmath,amssymb,amsfonts,english,jpgimgs]{revtex4}

\usepackage{babel}
\usepackage{graphicx}

\begin{document}

\title{Electromagnetic field near cosmic string\footnote{%
This paper is partially based on unpublished work \cite{Krtous:TPPKS}.}}

\author{Pavel Krtou\v{s}}
\email{Pavel.Krtous@mff.cuni.cz}

\affiliation{
  Institute of Theoretical Physics,
  Faculty of Mathematics and Physics, Charles University in Prague,\\
  V Hole\v{s}ovi\v{c}k\'{a}ch 2, 180 00 Prague 8, Czech Republic
  }

\date{June 6, 2006}

\begin{abstract}
The retarded Green function of the electromagnetic 
field in spacetime of a straight thin cosmic string is found.
It splits into a geodesic part (corresponding to the propagation along null rays) and
to the field scattered on the string. With help of the Green function 
the electric and magnetic fields of simple sources are constructed. 
It is shown that these sources are influenced 
by the cosmic string through a self-interaction with their field.
The distant field of static sources is studied and it is found that 
it has a different multipole structure than in Minkowski spacetime. 
On the other hand, the string suppresses the electric and magnetic field
of distant sources---the field is expelled from regions near the string. 
\end{abstract}

\pacs{11.27.+d, 03.65.Pm, 41.20.Cv, 03.50.De}


\maketitle



\section{Introduction}
\label{sc:intro}

One of the simplest mass sources in general relativity 
are \emph{cosmic strings}---linear objects with a given linear mass density and a linear tension.
Cosmic strings arise as topological defects in various gauge theories (see, e.g., \cite{Garfinkle:1985}),
or as a macroscopic variant of the fundamental strings (e.g., \cite{SarangiTye:2002}).
Thin cosmic strings can be phenomenologically described by conical singularities of 
the spacetime metric---one-dimensional \vague{objects} the angle around which is different than ${2\pi}$.
The static straight thin cosmic string is thus represented by 
a locally flat spacetime with a conical singularity along an axis. The deficit angle
is proportional to the linear mass density which is equal to the linear tension \cite{Vilenkin:1985}.
A string with a deficit angle has a positive mass density and it is stretched; 
a string with an excess angle has negative mass density and is squeezed.
For an overview of physics of cosmic strings see, e.g., \cite{VilenkinShellard:book}
and references therein; for more recent developments see~\cite{DavisKibble:2005}.

Beside the empty spacetime with a single string, cosmic strings appear in a wide variety
of solutions of Einstein equations. Any axially symmetric spacetime can be trivially 
modified to contain a string on the axis of symmetry. However, there exist also 
solutions where the string play a key physical role as, for example, C-metric.
Here the string is a physical agent causing the motion of black holes---see, e.g., 
\cite{KinnersleyWalker:1970,Bonnor:1982,Krtous:2005}. 
In a wide class of boost-rotation symmetric spacetimes \cite{BicakSchmidt:1989}
the strings accelerate even more general sources.

In the 80's the cosmic strings were considered as candidates 
for a mechanism of galaxy formation. This possibility was abandoned
mainly because of inconsistencies with cosmic microwave background observations.
Recently, however, the interest in cosmic strings reappeared 
in the context of the \vague{brane-world} scenarios
of the superstring theory. These suggest the existence of
macroscopic fundamental strings behaving 
as \vague{old-fashioned} cosmic strings 
(e.g., Refs.~\cite{DavisKibble:2005,Kibble:2004} and references therein). 
There were also tentative hints of
a detection of cosmic strings \cite{SazhinEtal:2003,SchildEtal:2004}
based on specific gravitational lensing effects,
but the explanation in terms of cosmic strings was not confirmed by 
subsequent observations \cite{AgolHoganPlotkin:2006}.

In the case of a single straight thin string the
curvature is localized only on the world-sheet of the string. 
Such spacetime thus represents a very simple but non-trival example which 
can serve as a toy model for studying various phenomena
due to curvature.

In this paper we investigate the behavior of the electromagnetic field
in the background of a non-charged cosmic string. In Sec.~\ref{sc:retgrfc} we find the retarded Green 
function and with its help we demonstrate the general behavior of the propagation
of the electromagnetic field \cite{DeWittBrehme:1960}: the field propagates 
(i) along null geodesics (on the light cone of the source),
and (ii) it is scattered on the curvature (so-called \vague{tail} term of the field).
In our case the field propagates on the deformed light cones of the source and 
it is scattered on the cosmic string.

In Sec.~\ref{sc:coulomb} we use the retarded Green function to derive 
the electric field of the static monopole and dipole sources. 
We obtain the field equivalent 
to that of Ref.~\cite{Linet:1986}, which was derived by a different method. 
We also recover that the source is self-interacting: a monopole charge is
repelled from the string by its own field, a dipole 
is forced to align \vague{around} the string.

A magnetic field of the current flowing along the line parallel with the string 
is derived in Sec.~\ref{sc:magfield}.

In Secs.~\ref{sc:farfield} and \ref{sc:homfield} we study 
the behavior of the electric field of strong static charges
at large distances and near the string. 
We find an interesting effect: the asymptotic field has a different monopole
structure from that in empty space, and the field of large charges 
which are localized far from the string is suppressed near the string.
The field which would be nonvanishing and homogeneous in an empty space 
is expelled from the region near the string. It means that the string \vague{shields}
its neighborhood from the influence of distant sources.

\section{The retarded Green function}
\label{sc:retgrfc}

Spacetime of a cosmic string is described by the flat metric ${\mtrc}$ with a
deficit angle around the string (see, e.g., \cite{VilenkinShellard:book})
\begin{equation}\label{metric}
\begin{gathered}
  \mtrc = -\grad t\,\grad t +\grad z\, \grad z
  +\grad\rho\,\grad\rho+\rho^2\grad\ph\,\grad\ph\commae\\
  t,z\in\realn\comma \rho\in\realn^+\comma \ph\in(-\pi/\iconpar,\pi/\iconpar)\period
\end{gathered}\end{equation}
The inverse conicity parameter ${\iconpar}$ characterizes the deficit angle,
${\Delta\ph=2\pi(1-\iconpar^{\!-1})}$.

Assuming the Lorentz  gauge condition, ${\covd_{\!\aix{n}}\EMA^{\aix{n}}=0}$,
the equation for the vector potential of the electromagnetic field in curved spacetime
outside the string (where ${\Ric=0}$) reduces to a simple wave equation
\begin{equation}\label{EMAeq}
\mtrc^{\!\aix{mn}}\covd_{\!\aix{m}}\covd_{\!\aix{n}}\EMA_{\!\aix{a}} = - \EMJ_{\!\aix{a}}\period
\end{equation}
To separate this vector equation to scalar equations it is useful to
project it onto a complex tetrad
\begin{equation}
\grad t\comma\!\!\grad z\comma\!\!
\mT=\isqrtwo\bigl(\grad\rho+i\rho\,\grad\ph\bigr)\comma\!\!
\bT=\isqrtwo\bigl(\grad\rho-i\rho\,\grad\ph\bigr)\commae
\end{equation}
in which the metric \eqref{metric} takes the form
\begin{equation}\label{metrictetr}
  \mtrc = -\grad t\,\grad t +\grad z\, \grad z
  +\mT\,\bT+\bT\,\mT\period
\end{equation}
The field equation \eqref{EMAeq} is then equivalent to scalar equations
\begin{subequations}\label{separation}
\begin{align}
\waveop[0]\EMAc_{t} &= - \EMJc_{t}\commae&
\waveop[0]\EMAc_{z} &= - \EMJc_{z}\comma\\
\waveop[+1]\EMAc_{\mTc} &= - \EMJc_{\mTc}\commae&
\waveop[-1]\EMAc_{\bTc} &= - \EMJc_{\bTc}\comma
\end{align}
\end{subequations}
where
\begin{subequations}\label{waveops}
\begin{equation}
\waveop[0]=\biggl[-\frac{\partial^2}{\partial t^2}+\frac{\partial^2}{\partial z^2}
  +\frac{\partial^2}{\partial \rho^2}+\frac{1}{\rho^2}\frac{\partial^2}{\partial \ph^2}
  +\frac{1}{\rho}\frac{\partial}{\partial \rho}\biggr]
\end{equation}
and
\begin{equation}
\waveop[\pm1]=\biggl[-\frac{\partial^2}{\partial t^2}+\frac{\partial^2}{\partial z^2}
  +\frac{\partial^2}{\partial \rho^2}+\frac{1}{\rho^2}\frac{\partial^2}{\partial \ph^2}
  +\frac{1}{\rho}\frac{\partial}{\partial \rho}
  \mp i\frac{2}{\rho^2}\frac{\partial}{\partial\ph}-\frac{1}{\rho^2}\biggr]\period
\end{equation}
\end{subequations}
The operator ${\waveop[0]}$ is the standard flat-space d'Alambert operator in cylindrical coordinates,
however, with coordinate ${\ph\in(-\pi/\iconpar,\pi/\iconpar)}$.

We look for eigenfuctions ${\efc{}(x)}$  of these operators in the form
${\efc{}(x)=f(\rho) \exp[-i(-\omega t +\kappa z + \iconpar n \ph)]}$
where ${t,z,\rho,\ph}$ are coordinates of a spacetime point ${x}$
and ${\omega,\kappa\in\realn}$, ${n\in\integern}$ are parameters
labeling the eigenfunctions. The restriction on ${n}$ follows from
the periodicity of the angular coordinate for ${\ph=\pm\pi/\iconpar}$.
We find that ${f(\rho)}$ must satisfy  Bessel equation
\begin{equation}  
  \biggl[\frac{\partial^2}{\partial \rho^2}+\frac{1}{\rho}\frac{\partial}{\partial \rho}
  +\Bigl(\lambda^2-\frac{(\iconpar n+\sg)^2}{\rho^2}\Bigr)\biggr]f=0
\end{equation}
for some positive number ${\lambda\in\realn^+}$.
The complete system of solutions of  Bessel equation regular 
on the string (i.e., for ${\rho=0}$) is given by Bessel functions ${\BesJ{\alpha}}$ with ${\alpha>0}$.
The eigenfunctions of the operators \eqref{waveops} thus are
\begin{equation}\label{efc}
\efc{\omega,\kappa,\lambda,n}(x)
  ={\textstyle\frac{\iconpar^{1/2}}{(2\pi)^{3/2}}}\;
  e^{-i(-\omega t +\kappa z + \iconpar n \ph)}
  \BesJ{\abs{\iconpar n+\sg}}(\lambda\rho)
  \period
\end{equation} 
They satisfy
\begin{equation}  
  \waveop\,\efc{\omega,\kappa,\lambda,n}=-(-\omega^2+\kappa^2+\lambda^2)\;\efc{\omega,\kappa,\lambda,n}
\end{equation}
and the normalization was chosen in such a way that the following completeness and orthonormality relations hold:
\begin{equation}
\begin{split}
  &\sum_{\omega,\kappa,\lambda,n}\efc{\omega,\kappa,\lambda,n}(x)\; \befc{\omega,\kappa,\lambda,n}(x') \\
    &\quad\equiv\int_\realn\! d\omega\int_\realn\! d\kappa\int_{\realn^+}\!\!\!\!\lambda\, d\lambda\sum_{n\in\integern} 
    \efc{\omega,\kappa,\lambda,n}(x)\, \befc{\omega,\kappa,\lambda,n}(x')\\
    &\quad=\delta(x|x')\commae
\end{split}
\end{equation}
\begin{equation}
\begin{split}
  &\bigl(\efc{\omega,\kappa,\lambda,n}\,, \efc{\omega',\kappa',\lambda',n'}\bigr) \\
    &\quad\equiv\int_\realn\! d t\!\int_\realn\! d z\!
      \int_{\realn^+}\!\!\!\!\rho\, d\rho\! \int_{\!\!-\!\pi\!/\!\iconpar}^{\pi\!/\!\iconpar} \!\!\!\!\!\!d\ph\; 
    \efc{\omega,\kappa,\lambda,n}(x)\, \befc{\omega',\kappa',\lambda',n'}(x)\\
    &\quad=\delta(\omega,\kappa,\lambda|\omega',\kappa',\lambda')\,\delta_{nn'}\period
\end{split}
\end{equation}
The spacetime delta-function ${\delta(x|x')}$ is normalized with respect to the measure ${\mtrcvol=\rho\, dt\, dz\, d\rho\, d\ph}$,
the delta-function in space of parameters ${\delta(\omega,\kappa,\lambda|\omega',\kappa',\lambda')}$
is normalized using ${\lambda\, d\omega\, d\kappa\, d\lambda}$, and ${\delta_{nn'}}$ is Kronecker delta.

The retarded Green function ${\grfc(x|x')}$ of the operator~${\waveop}$,
\begin{subequations}  \label{retgrfccond}
\begin{gather}
  \waveop_x\,\grfc(x|x') = \delta(x|x')\comma\\
  \supp_x\grfc(x|x')\subset \text{future}(x')\comma \label{retcond}
\end{gather}
\end{subequations}
can be decomposed in the system \eqref{efc} as follows:
\begin{equation}\label{grfcint}
\begin{split}
&\grfc(x|x')=\sum_{\omega,\kappa,\lambda,n}{}_{\!\!\!\!\!\raisebox{-4pt}{$\scriptstyle\ret$}\;}
   \frac{\efc{\omega,\kappa,\lambda,n}(x)\, \befc{\omega,\kappa,\lambda,n}(x')}{-\omega^2+\kappa^2+\lambda^2}\\
&\quad\equiv\frac{\iconpar}{(2\pi)^3}
   \sum_{n\in\integern} \int_{\realn^+}\!\!\!\!\lambda\, d\lambda\!\int_\realn\!d\kappa\!\int_{c_\ret}\!\!\!\!\!\! d\omega\,
   \frac{e^{-i(-\omega\Delta t+\kappa\Delta z+\iconpar n \Delta\ph)}}{-\omega^2+\kappa^2+\lambda^2}\\
   &\mspace{220mu}\times\BesJ{\abs{\iconpar n+\sg}}(\lambda\rho)\BesJ{\abs{\iconpar n+\sg}}(\lambda\rho')\commae
\end{split}\raisetag{16.5ex}
\end{equation}
where ${\Delta t = t-t'}$, ${\Delta z = z-z'}$, ${\Delta\ph = \ph-\ph'}$, and
${c_\ret}$ is the path along the real axis in the complex plane of parameter ${\omega}$.
It goes around the poles at ${\omega=\pm\sqrt{\kappa^2+\lambda^2}}$ in lower half plane 
(${\im\omega<0}$) to satisfy the conditions \eqref{retcond}.

A nontrivial integration performed in the Appendix leads to the expression
\begin{widetext}
\begin{equation}  \label{grfcexpl}
\begin{split}
  &\grfc(x|x') = \frac1{2\pi}\,\theta(\Delta t)\sum_{k=\kin}^{\kfi}e^{i\sg(\Delta\ph+2\pi k/\iconpar)}\,\delta\bigl(-\Delta t^2+r_k^2\bigr)
    -(-1)^\sg\frac{\iconpar}{8\pi^2}\frac{\theta(\Delta t)\,\theta\bigl(\Delta t^2\!-\!\Delta z^2\!-\!(\rho\!+\!\rho')^2\bigr)}{\rho\rho'}\\
    &\quad\times\biggl[\frac{\ch\sg\eta}{\sh\eta}\Bigl(
    \frac{\sin\iconpar(\pi-\Delta\ph)}{\ch\iconpar\eta\!-\!\cos\iconpar(\pi\!-\!\Delta\ph)}
    +\frac{\sin\iconpar(\pi+\Delta\ph)}{\ch\iconpar\eta\!-\!\cos\iconpar(\pi\!+\!\Delta\ph)}\Bigr)
    +i\sg\Bigl(\frac{\sh\iconpar\eta}{\ch\iconpar\eta\!-\!\cos\iconpar(\pi\!-\!\Delta\ph)}
    -\frac{\sh\iconpar\eta}{\ch\iconpar\eta\!-\!\cos\iconpar(\pi\!+\!\Delta\ph)}\Bigr)\biggr]\commae
\end{split}\raisetag{12ex}
\end{equation}
\end{widetext}\pagebreak
\begin{figure}[t]
\includegraphics{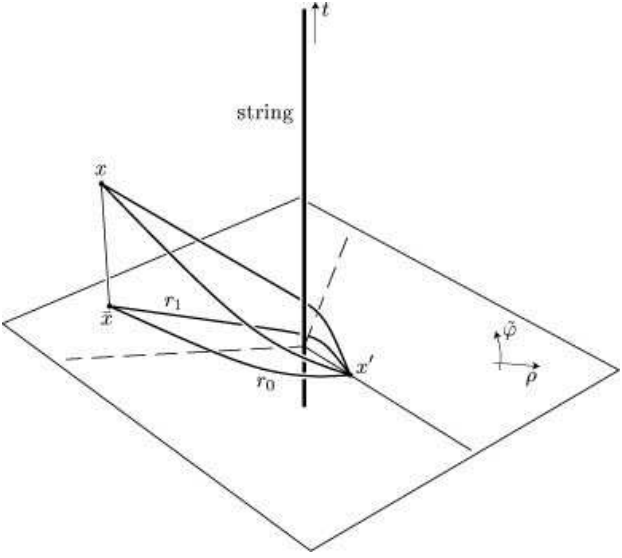}
\\
\caption{\label{fig:geod}%
Three-dimensional spacetime diagram of the plane orthogonal to the cosmic string
(i.e., the spacetime diagram with coordinate ${z}$ is suppressed). 
A rescaled coordinate ${\tph=\iconpar\ph}$ (periodic with the period ${2\pi}$)
is used instead of \vague{geometrical} angular coordinate ${\ph\in(-\pi/\iconpar,\pi/\iconpar)}$. Two spacetime points ${x}$
and ${x'}$ can be joined by more than one geodesic as indicated in the diagram.
The geodesics \vague{bend} around the string due to the curvature localized on the string.
Of course, the \vague{bending} of the geodesics is only apparent---it 
arises because the rescaled coordinate ${\tph}$ is employed.
The geodesics are straight lines in locally flat geometry. 
However, the cosmic string causes the angle deficit and the intersection
of geodesics behind the string is a real effect. Spatial projection of the geodesics
into hypersurface ${t=\text{constant}}$ is also shown. These are spatial geodesics 
with length ${r_k}$. Also cf.\ Fig.~\ref{fig:spatial}.%
}%
\end{figure}
where ${\eta}$ and ${r_k}$ are defined as
\begin{align}
  &\ch\eta=\frac{\Delta t^2-\Delta z^2-\rho^2-\rho'^2}{2\rho\rho'}\commae\label{etadef}\\
  &r_k^2=\Delta z^2+\rho^2+\rho'^2-2\rho\rho'\cos\bigl(\Delta\ph+2\pi k/\iconpar\bigr)\commae\label{rkdef}
\end{align}
and ${\theta(t)}$ is the Heaviside step function.
As we will discuss in detail below, index ${k\in\integern}$ 
labels spacetime geodesics joining points ${x}$ and ${x'}$;
its range ${\kin(\Delta\ph),\dots,\kfi(\Delta\ph)}$ is given by conditions\footnote{%
For brevity, we will not write the dependence of 
${\kin}$ and ${\kfi}$ on angle ${\Delta\ph}$ explicitly.}
\begin{equation}\label{kinfi}
\begin{gathered}
\Delta\ph+2\pi(\kin-1)/\iconpar<-\pi<\Delta\ph+2\pi\kin/\iconpar\commae\\
\Delta\ph+2\pi\kfi/\iconpar<\pi<\Delta\ph+2\pi(\kfi+1)/\iconpar\period
\end{gathered}
\end{equation}

\begin{figure}[b]
\includegraphics{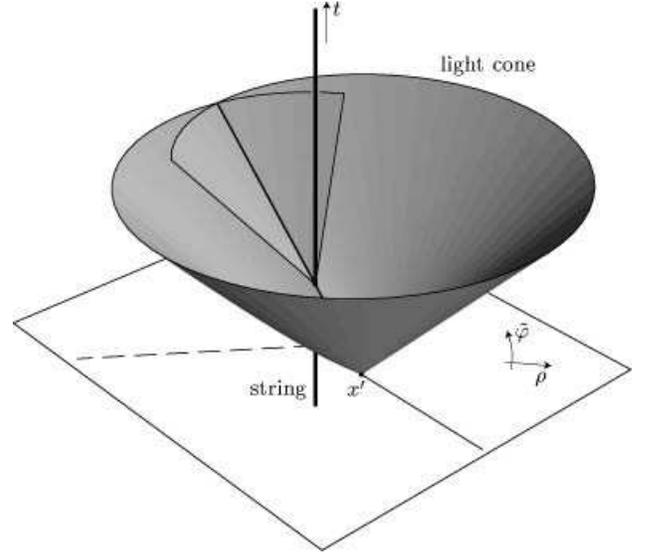}
\\
\caption{\label{fig:cone}%
The support of the geodesic part ${\vgrfc[\geod](x|x')}$ of the Green
function is localized on the future light cone of the source point ${x'}$,
i.e., on the null hypersurface generated by future null geodesics from ${x'}$.
Near vertex ${x'}$ the hypersurface has the standard structure of 
the light cone in Minkowski spacetime. However, at the cosmic string the light cone 
deforms---it intersects itself and becomes a hypersurface with a boundary
(given by null rays propagating near the string).%
}%
\end{figure}

In \eqref{separation} we separated the field equations into independent ones for components ${\EMAc_t,\EMAc_z,\EMAc_\mTc,\EMAc_\bTc}$.
Now we can combine the Green functions ${\grfc(x|x')}$ 
back to the full vector Green function:
\begin{equation}\label{vgrfc}
\begin{split}
\vgrfc_{\aix{ab}'}(x|x')&=
  \bigl(-\,\grad_{\aix{a}}t\,\grad_{\aix{b}'}t+\grad_{\aix{a}}z\,\grad_{\aix{b}'}z\bigr)\grfc[0](x|x')\\
  &\;\;+ \mT_{\aix{a}}\,\bT_{\aix{b}'}\grfc[+1](x|x')+ \bT_{\aix{a}}\,\mT_{\aix{b}'}\grfc[-1](x|x')\commae
\end{split}\raisetag{7ex}
\end{equation}
with unprimed and primed tensor indices considered at point ${x}$ and ${x'}$, respectively.
As a consequence of \eqref{separation} and \eqref{retgrfccond} the vector Green function
satisfies the conditions analogous to \eqref{retgrfccond} with the wave operator from Eq.~\eqref{EMAeq}.

This Green function can be split into two pieces with
clear interpretation,
\begin{equation}\label{grfcsplit}
\vgrfc_{\aix{ab}'}(x|x')=\vgrfc[\geod]_{\aix{ab}'}(x|x')+\vgrfc[\scat]_{\aix{ab}'}(x|x')\period
\end{equation}
The first \emph{geodesic} part
\begin{equation}\label{grfcgeod}
\vgrfc[\geod]_{\aix{ab}'}(x|x')=
   \frac1{2\pi}\,\theta(\Delta t)\sum_{k=\kin}^{\kfi} \geodtr_{\aix{ab}'}[\gamma_k]\,\delta\bigl(-\Delta t^2+r_k^2\bigr)
\end{equation}
describes the propagation of the electromagnetic field along the null rays as in an empty flat spacetime.
Here ${\gamma_k}$, ${k=\kin,\dots,\kfi}$ are geodesics joining points ${x}$ and ${x'}$, and
${\geodtr_{\aix{ab}'}[\gamma]}$ is the operator of parallel transport along the geodesic ${\gamma}$.
Quantity ${r_k}$ defined in \eqref{rkdef} is the spatial length of the ${k}$-th geodesic, see Fig.~\ref{fig:geod}.
Delta-functions in \eqref{grfcgeod} enforce that the field propagates only along null geodesics.
The only difference from the Minkowski spacetime is that light rays \vague{bend} around the cosmic string.
If we call \emph{light cone} of the source point ${x'}$ a hypersurface generated by null geodesics from ${x'}$
we see that it \vague{deforms} when it intersects the string -- see Fig.~\ref{fig:cone}.
The contribution to the field with the source at point ${x'}$
described by the geodesic part \eqref{grfcgeod}
is fully localized on this light cone.

\begin{figure}
\includegraphics{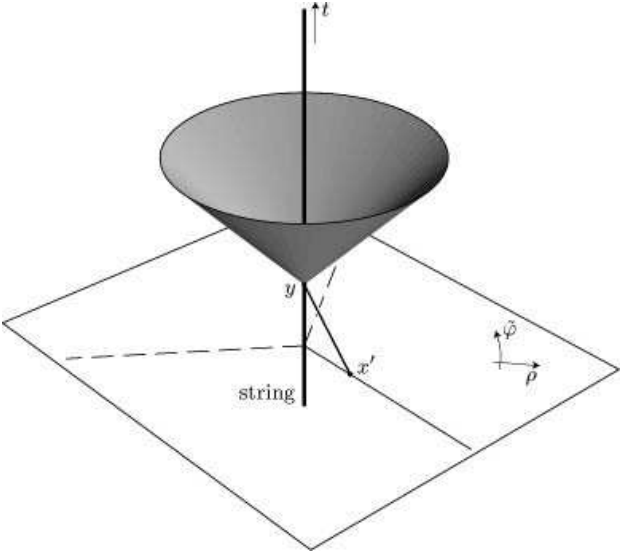}
\\
\caption{\label{fig:scat}%
The support of the scattered part ${\vgrfc[\scat](x|x')}$ of the Green
function is localized inside light cones with vertices ${y}$
on the cosmic string which are connected with the source by null geodesics. 
In the figure only a spacetime diagram of the plane going through the charge orthogonally 
to the string is shown.
The electromagnetic field given by ${\vgrfc[\scat](x|x')}$ can be interpreted as the electromagnetic field 
scattered on the singular curvature localized on the string.%
}%
\end{figure}

If we transform 1-forms ${\mT,\bT}$
back to the coordinate \mbox{1-forms} ${\grad\rho}$ and ${\grad\ph}$, 
the second part, ${\vgrfc[\scat](x|x')}$, of the Green function becomes
\begin{widetext}
\begin{equation}\label{grfcscat}
\begin{split}
  &\vgrfc[\scat]_{\aix{ab}'}(x|x') =
    -\frac{\iconpar}{8\pi^2}\frac{\theta(\Delta t)\,\theta\bigl(\Delta t^2\!-\!\Delta z^2\!-\!(\rho\!+\!\rho')^2\bigr)}{\rho\rho'}\\
    &\quad\times\biggl[\biggl(\!\bigl(-\,\grad_{\aix{a}}t\,\grad_{\aix{b}'}t+\grad_{\aix{a}}z\,\grad_{\aix{b}'}z\bigr)
    \!\!-\!\bigl(\grad_{\aix{a}}\rho\,\grad_{\aix{b}'}\rho+\rho\rho'\grad_{\aix{a}}\ph\,\grad_{\aix{b}'}\ph\bigr)\ch\eta\biggr)
    \frac{1}{\sh\eta}\biggl(
    \frac{\sin\iconpar(\pi-\Delta\ph)}{\ch\iconpar\eta\!-\!\cos\iconpar(\pi\!-\!\Delta\ph)}
    +\!\frac{\sin\iconpar(\pi+\Delta\ph)}{\ch\iconpar\eta\!-\!\cos\iconpar(\pi\!+\!\Delta\ph)}\biggr)\\
    &\mspace{50mu}+\bigl(\rho\,\grad_{\aix{a}}\ph\,\grad_{\aix{b}'}\rho-\rho'\,\grad_{\aix{a}}\rho\,\grad_{\aix{b}'}\ph\bigr)\,
    \biggl(\frac{\sh\iconpar\eta}{\ch\iconpar\eta\!-\!\cos\iconpar(\pi\!-\!\Delta\ph)}
    -\frac{\sh\iconpar\eta}{\ch\iconpar\eta\!-\!\cos\iconpar(\pi\!+\!\Delta\ph)}\biggr)\biggr]\period
\end{split}\raisetag{18ex}
\end{equation}
\end{widetext}
It can be associated with a \emph{scattering} on the 
\vague{curvature} localized
on the string. Indeed, the contribution to the field
due to this part of the Green function is
localized in the causal future of points ${y}$ on the string 
which are connected to the source point ${x'}$ by null geodesics
-- see Fig.~\ref{fig:scat}. This fact is a simple manifestation
of general behavior of the propagation of the electromagnetic field
in a curved spacetime \cite{DeWittBrehme:1960}: the main part propagates along null rays
and is supported on the light cone of the source point. However,
the field is scattered by curvature and propagates
also inside light cone of the source point. This is the \vague{tail-part}
of the field. It is easily seen that ${\vgrfc[\scat](x|x')}$ vanishes for ${\iconpar=1}$
(Minkowski space). The analogous behavior was also found in 
Refs.~\cite{Khusnutdinov:1995,AlievGaltsov:1989} where
the field of a general pointlike source was discussed.

\section{Electric field of the static charge and dipole}
\label{sc:coulomb}

\begin{figure*}
\includegraphics{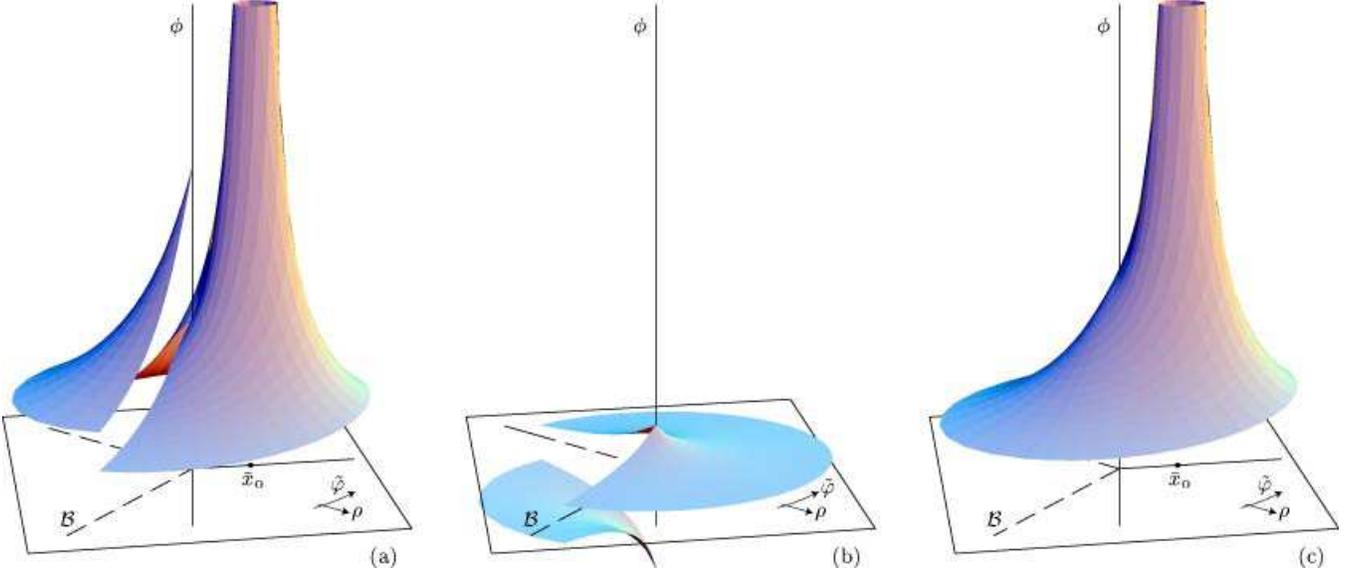}
\caption{\label{fig:scpot}%
The graph of the scalar potential of a static charge near a cosmic string. Horizontal plane
corresponds to the plane of the charge orthogonal to the cosmic string.
The values of the scalar potential are drawn in the vertical direction.
The full potential \eqref{scpot} (diagram (c)) is split into
the geodesic part (a) and the scattered part (b). The whole potential
is smooth except at the locations of the charge and at the string.
Both geodesic and scattered parts are discontinuous
at surfaces ${\mathcal{B}}$ where a number of spatial geodesics from the charge changes 
(cf.\ Fig.~\ref{fig:spatial}), however, the discontinuity 
of these two parts of the Green function cancels each other out.
The scattered part of the field is finite and smooth near the charge; 
this part is responsible for self-interaction of the charge.%
}%
\vspace*{-2ex}
\end{figure*}

The Green function \eqref{vgrfc} can be used to derive
the electric field of the static charge.
Let us consider charge ${\charge}$ at
${z=z_\oix}$, ${\rho=\rho_\oix}$, and ${\ph=\ph_\oix}$. 
The integration of the Green function over the electric current ${\EMJ}$ 
simplifies: the source is static and only the time component
of the vector potential survives; the integration over delta-functions of the  geodesic part 
reduces to a simple sum; and thanks to the step functions in \eqref{grfcscat} 
the scattered part leads to the integration in proper time ${\tau}$ over half line only. 
Introducing the scalar potential ${\EMP(x) = -\EMAc_t(x)}$ we get
\begin{equation}\label{scpot}  
\begin{split}
&\EMP(x) = \frac{\charge}{4\pi}\sum_{k=\kin}^{\kfi}\frac1{r_k}\\
  &\quad-\frac{\iconpar\charge}{8\pi^2}\int_{0}^{\infty}
  \frac{\sfc(\eta,\Delta\ph)\,d\eta}{\sqrt{\Delta z^2+\rho^2+\rho_\oix^2+2\rho\rho_\oix\ch\eta}}
\period
\end{split}
\end{equation}
Here we defined function ${\sfc}$ by
\begin{equation}\label{Sdef}
  \sfc(\eta,\ph) = 
  \frac{\sin\iconpar(\pi-\ph)}{\ch\iconpar\eta\!-\!\cos\iconpar(\pi\!-\!\ph)}
  +\frac{\sin\iconpar(\pi+\ph)}{\ch\iconpar\eta\!-\!\cos\iconpar(\pi\!+\!\ph)}\commae
\end{equation}
and we used notation
\begin{equation}  
\begin{aligned}
  &\Delta z = z-z_\oix\comma \Delta\ph=\ph-\ph_\oix\commae\\
  &r_k^2=\Delta z^2+\rho^2+\rho_\oix^2-2\rho\rho_\oix\cos\bigl(\Delta\ph+2\pi k/\iconpar\bigr)\period
\end{aligned}
\end{equation}
The first term---the sum in \eqref{scpot}---has the origin 
in the geodesic part of the Green function,
the integral term arises from the scattered part. 
In derivation of this term we changed the integration 
over proper time ${\tau}$ into the integration over ${\eta}$ using the substitution 
\begin{equation}  
  \ch\eta=\frac{(t-\tau)^2-\Delta z^2-\rho^2-\rho_\oix^2}{2\rho\rho_\oix}\period
\end{equation}
For a graphical representation of the scalar potential, see Fig.~\ref{fig:scpot}.

The electric field ${\EME=\EMEc^{\hat\rho} \tens{e}_\rho+\EMEc^{\hat\ph} \tens{e}_\ph+\EMEc^{\hat z} \tens{e}_z}$ 
implied by \eqref{scpot}, evaluated with respect to the normalized triad ${\{\tens{e}_\rho,\,\tens{e}_\ph,\,\tens{e}_z\}}$
(i.e., ${\tens{e}_\rho=\cvil{\rho}}$, ${\tens{e}_\ph=\rho^{-1}\cvil{\ph}}$, ${\tens{e}_z=\cvil{z}}$) is
\begin{equation*}
\begin{split}
  \EMEc^{\hat z}&=\frac{\charge}{4\pi}\sum_{k=\kin}^{\kfi}\frac{\Delta z}{r_k^3}\\
         &\quad-\frac{\iconpar\charge}{8\pi^2}\int_0^{\infty}
         \frac{\sfc(\eta,\Delta\ph)\,\Delta z\,d\eta}{\bigl(\Delta z^2+\rho^2+\rho_\oix^2+2\rho\rho_\oix\ch\eta\bigr)^{3/2}}\commae
\end{split}\raisetag{10ex}
\end{equation*}
\begin{equation}\label{elf}
\begin{split}
  \EMEc^{\hat\rho}&=\frac{\charge}{4\pi}\sum_{k=\kin}^{\kfi}\frac{\rho-\rho_\oix\cos(\Delta\ph+2\pi k/\iconpar)}{r_k^3}\\
         &\quad-\frac{\iconpar\charge}{8\pi^2}\int_0^{\infty}
         \frac{\sfc(\eta,\Delta\ph)\,(\rho+\rho_\oix\ch\eta)\,d\eta}{\bigl(\Delta z^2+\rho^2+\rho_\oix^2+2\rho\rho_\oix\ch\eta\bigr)^{3/2}}\commae
\end{split}\raisetag{10ex}
\end{equation}
\begin{equation*}
\begin{split}
  \EMEc^{\hat\ph}&=\frac{\charge}{4\pi}\sum_{k=\kin}^{\kfi}\frac{\rho_\oix\sin(\Delta\ph+2\pi k/\iconpar)}{r_k^3}\\
         &\quad+\frac{\iconpar\charge}{8\pi^2}\int_0^{\infty}
         \frac{\frac{\partial\sfc}{\partial\ph}(\eta,\Delta\ph)\,d\eta}{\rho\bigl(\Delta z^2+\rho^2+\rho_\oix^2+2\rho\rho_\oix\ch\eta\bigr)^{1/2}}\period
\end{split}\raisetag{10ex}
\end{equation*}

\begin{figure}
\vspace*{-3ex}
\includegraphics{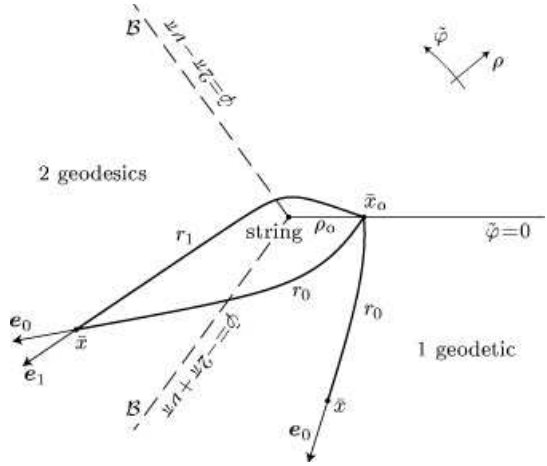}
\\
\caption{\label{fig:spatial}%
The spatial diagram of the hypersurface ${t=\text{constant}}$ (cf.\ spacetime diagram in Fig.~\ref{fig:geod}). 
Points ${\bar x}$ and ${\bar x_\oix}$ are spatial projections of spacetime point ${x}$ and of the 
worldline of the static charge. Rescaled coordinate ${\tilde\ph=\iconpar\ph}$ is used.
geodesic part \eqref{elfgeod} of the electric field at ${\bar x}$ of 
the charge at ${\bar x_\oix}$ is given 
by standard flat-space Coulomb contributions 
for each spatial geodesic joining ${\bar x}$ and ${\bar x_\oix}$.
Surface ${\mathcal{B}}$ (indicated by dashed line) 
divides the space into domains, points of which are connected 
with the charge by one geodesic or by two geodesics, respectively. 
For sufficiently large angle deficit there exist also points
connected with the charge by more geodesics.
}%
\vspace*{-2ex}
\end{figure}

Geodesic part has a clear meaning---it is a sum of the \vague{Coulomb} terms corresponding to
different spatial geodesics joining the charge and the field point. Indeed, it has the form
\begin{equation}\label{elfgeod}
\EME^{\geod}(x)=\sum_{k=\kin}^{\kfi}\frac{\charge}{4\pi}\,\frac{1}{r_k^2}\,\tens{e}_k\commae
\end{equation}
where, ${\tens{e}_k}$ is a unit vector tangent to the ${k}$-th spatial geodesic 
joining ${\bar x_\oix}$ and ${\bar x}$---the spatial projections (projections to the hypersurface 
${t=\text{constant}}$) of the charge and of the field point ${x}$,
cf.\ Fig.~\ref{fig:spatial}.
Clearly, the geodesic part of the field is discontinuous on surfaces
on one side of which points are connected to the charge 
by different number of geodesics than on other side.
However, the whole electric field \eqref{elf}
is continuous here. The discontinuity is compensated 
by the scattered part of the field; cf.\ the graphs for the scalar potential in Fig.~\ref{fig:scpot}.

Near the cosmic string we can observe an interesting phenomenon:  
the charge is self-interacting with itself. 
Of course, the field evaluated at the charge is infinite.
However, if we subtract the standard Coulomb field near the 
charge, we obtain a finite residual field which acts on the charge itself.
The regularized potential energy of the charge in its field is equal to
\begin{equation} \label{monselfenrg}
W^\slf(\bar x_\oix)=\frac\charge2 \Bigl(\EMP-\frac{\charge}{4\pi}\frac1{r_0}\Bigr)\Big\vert_{x_\oix}
 =C\frac{\charge^2}{\rho_\oix}\commae
\end{equation}
with the constant ${C}$ given by\footnote{%
${\lfloor\iconpar/2\rfloor}$ denotes an integer part of ${\iconpar/2}$. 
Indeed, conditions \eqref{kinfi} for ${\Delta\ph=0}$ give ${\kfi=-\kin=\lfloor\iconpar/2\rfloor}$,
and ${k=0}$ term is removed by the regularization.
For ${\iconpar<2}$ the sum contains thus no terms.}
\begin{equation}\label{Cconst}
C=\frac{1}{8\pi}\sum_{k=1}^{\lfloor\iconpar/2\rfloor}\frac1{\sin\frac{\pi k}\iconpar}
-\frac{\iconpar\,\sin\iconpar\pi}{16\,\pi^2}
\int_0^\infty\!\!\!\!\frac{\ch^{\!-\!1}\!\frac\eta2}{\ch\iconpar\eta-\cos\iconpar\pi}\,d\eta\period
\end{equation}
The self-force is
\begin{equation}\label{elselfforce}  
\tens{F}^{\,\slf}= C \frac{\charge^2}{\rho_\oix^2}\;\tens{e}_{\rho}\period
\end{equation}

The result \eqref{scpot} and the prediction of the self-interaction are not new.
They are equivalent to the result of Ref.~\cite{Linet:1986} which was
obtained by a direct solution of the three-dimensional Laplace equation in the space with angle deficit.
Nevertheless, our form is more useful for the calculation of self-force \eqref{elselfforce}
since we can easily subtract the divergent flat space contribution which correspond to the ${k=0}$
term in sums in \eqref{scpot} and \eqref{elf}. The self-interaction of a general source 
was also discussed in Refs.~\cite{Khusnutdinov:1995,AlievGaltsov:1989}.

Using the electric field of the static point charge it is straightforward to 
find the field of an electric dipole at ${\bar x_\oix}$ with the spatial charge distribution given by
${-\tens{p}^{\aix{n}}\covd_{\aix{n}}\delta(\bar x|\bar x_\oix)}$.
For ${\tens{p}=p^{\hat z}\tens{e}_z+p^{\hat \rho}\tens{e}_\rho+p^{\hat \ph}\tens{e}_\ph}$ the potential is
\begin{equation}\label{eldip}
\begin{split}
  \EMP=&\,\frac{1}{4\pi}\sum_{k=\kin}^\kfi\frac1{r_k^3}
  \Bigr[\Delta z\, p^{\hat z}-\bigl(\rho_\oix-\rho\cos\Delta\ph_k\bigr) p^{\hat\rho}\\[-3ex]
  &\mspace{240mu}+\rho\sin\Delta\ph_k\, p^{\hat\ph}\Bigr]\\[1ex]
  &+\frac\iconpar{8\pi^2}\int_0^\infty\biggl[\frac{\sfc(\eta,\Delta\ph)
  \bigl(-\Delta z\, p^{\hat z}+(\rho_\oix\!+\!\rho\ch\eta)p^{\hat\rho}\bigr)}
  {\bigl(\Delta z^2+\rho^2+\rho_\oix^2+2\rho\rho_\oix\ch\eta\bigr)^{3/2}}\\
  &\mspace{100mu}+\frac{\rho_\oix^{-1}\;\frac{\partial \sfc}{\partial\ph}(\eta,\Delta\ph)\,p^{\hat\ph}}
  {\bigl(\Delta z^2+\rho^2+\rho_\oix^2+2\rho\rho_\oix\ch\eta\bigr)^{1/2}}\biggr]\,d\eta\commae
\end{split}
\end{equation}
where ${\Delta\ph_k=\Delta\ph+2\pi k/\iconpar}$.

The action of the dipole on itself can be obtained from the (regularized)
self-energy of the dipole in its own field defined analogously to \eqref{monselfenrg}.
Calculations lead to
\begin{equation}\label{dipselfenrg}
  W^\slf(\bar x_\oix,\tens{p}) = \frac1{\rho_\oix^3}\,
  \Bigl( C_{(z)}(p^{\hat z})^2+C_{(\rho)}(p^{\hat\rho})^2-C_{(\ph)}(p^{\hat\ph})^2\Bigr)\commae
\end{equation}
where positive\footnote{%
For ${\iconpar<2}$, the positivity of ${C_{(\rho)}}$ and ${C_{(z)}}$ follows 
immediately from expressions \eqref{Cdipconst}. The positivity for ${\iconpar>2}$ 
and the positivity of ${C_{(\ph)}}$ was checked numerically.}
constants ${C_{(\rho)}}$, ${C_{(z)}}$, and ${C_{(\ph)}}$
are given by 
\begin{align}
C_{(z)}\!&=\frac{1}{16\pi}\!\!\sum_{k=1}^{\lfloor\iconpar/2\rfloor}\!\frac1{\sin^3\frac{\pi k}\iconpar}
-\frac{\iconpar\,\sin\iconpar\pi}{32\,\pi^2}\!\!
\int_0^\infty\!\!\!\!\frac{\ch^{\!-\!3}\!\frac\eta2}{\ch\iconpar\eta\!-\!\cos\iconpar\pi}\,d\eta\commae\notag\allowdisplaybreaks\\
\begin{split}
C_{(\rho)}\!&=\frac{1}{32\pi}\!\sum_{k=1}^{\lfloor\iconpar/2\rfloor}\frac{3-\cos\frac{2\pi k}{\iconpar}}{\sin^3\frac{\pi k}\iconpar}\\
            &\;\quad-\frac{\iconpar\,\sin\iconpar\pi}{64\,\pi^2}
            \int_0^\infty\!\!\!\frac1{\ch^{3}\!\frac\eta2}\,\frac{3+\ch\eta}{\ch\iconpar\eta\!-\!\cos\iconpar\pi}\,d\eta\commae
\end{split}\label{Cdipconst}\allowdisplaybreaks\\
\begin{split}
C_{(\ph)}\!&=\frac{1}{32\pi}\!\sum_{k=1}^{\lfloor\iconpar/2\rfloor}\frac{3+\cos\frac{2\pi k}{\iconpar}}{\sin^3\frac{\pi k}\iconpar}\\
            &\;\quad+\frac{\iconpar^3\,\sin\iconpar\pi}{8\,\pi^2}
            \int_0^\infty\!\!\!\frac1{\ch\!\frac\eta2}\,
            \frac{\ch^2\!\iconpar\eta\!+\!\ch\iconpar\eta\cos\iconpar\pi\!-\!2}{(\ch\iconpar\eta\!-\!\cos\iconpar\pi)^3}\,d\eta\period
\end{split}\notag
\end{align}
The dipole is acting on itself by force
\begin{equation}\label{edipselfforce}
  \tens{F}^{\,\slf}=-\bar\covd W^\slf(\bar x_\oix,\tens{p})
\end{equation}
and by a torque 
\begin{equation}\label{edipselftorque}
  \tens{\tau}^{\,\slf}=-\tens{p}\times\frac{\partial W^\slf}{\partial \tens{p}}(\bar x_\oix,\tens{p})\commae
\end{equation}
where ${\bar\covd}$ is spatial covariant derivative and ${\times}$ spatial cross-product.

Observing the structure of the self-energy \eqref{dipselfenrg} we see that 
the torque is vanishing if the dipole is parallel to directions ${\tens{e}_\rho}$,
${\tens{e}_z}$, or ${\tens{e}_\ph}$; the equilibrium 
is stable for direction ${\tens{e}_\ph}$. The self-force
is repulsive from the string for the dipole along 
${\tens{e}_\rho}$ and ${\tens{e}_z}$ directions, and it is attractive 
towards the string for the dipole along ${\tens{e}_\ph}$ direction.

\section{Magnetic field of the current parallel to the string}
\label{sc:magfield}

\begin{figure}
\includegraphics{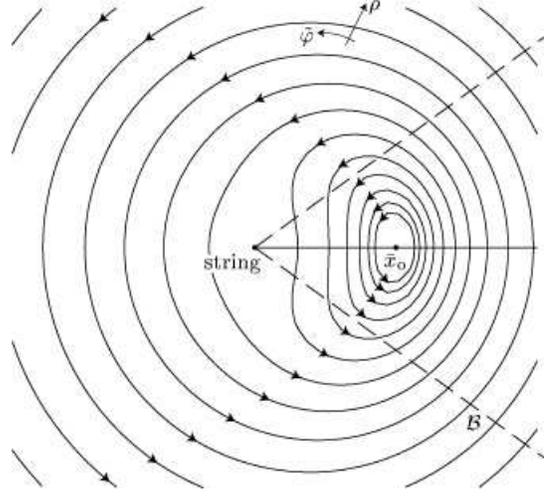}
\\
\caption{\label{fig:maglines}%
Magnetic lines around the electric current along the 
line through ${\bar x_\oix}$ parallel to the string.
Planes ${\mathcal{B}}$ (indicated by dashed lines) separate 
the space into domains in which the points are connected 
with the source by one, respectively two geodesics. 
A rather large value ${\iconpar=1.8}$ of the inverse conicity parameter 
was chosen to emphasize the deformation of magnetic lines
near the string. As discussed in Sec.~\ref{sc:homfield}, 
for physically relevant values, ${\iconpar\approx 1}$,
the domain near the string in which the magnetic 
lines are deformed is too small to be visualized
(cf.\ also Fig.~\ref{fig:fltdomsize}).}%
\vspace*{-2ex}
\end{figure}

Analogous results can be obtained also in the case
of only two relevant spatial dimensions, i.e., 
with the sources distributed homogeneously
along the string. The simplest interesting example is 
the field of the static electric current ${I}$ flowing in the ${z}$ direction along the line  
at a distance ${\rho_\oix}$ from the string. It follows from the Green function
\eqref{vgrfc} that for the current in the ${z}$ direction
the only nonvanishing component of the vector potential is ${\EMAc_{z}}$, 
and it is given by the same scalar Green function as in the electrostatic case.
Integrating the Green function over 
the world-sheet of the line splits into the integration over 
the time direction---which is equivalent to the integration performed 
in the previous section---and to the integration over the ${z}$ direction.
The latter integration diverges because of an infinite length
of the source, however, this divergence can be removed 
shifting the potential by an infinite constant\footnote{%
This divergence is not related to the string---the same situation occurs also 
in empty Minkowski spacetime.}.
For the ${z}$ component of the vector potential we obtain
\begin{equation}\label{vecpot}  
\begin{split}
&\EMAc_{z}(x) = -\frac{I}{2\pi}\sum_{k=\kin}^{\kfi}\ln{s_k}\\
  &\quad+\frac{\iconpar I}{8\pi^2}\int_{0}^{\infty}
  \sfc(\eta,\Delta\ph)\ln\bigl({\rho^2+\rho_\oix^2+2\rho\rho_\oix\ch\eta}\bigr)\,d\eta
\commae
\end{split}\raisetag{9ex}
\end{equation}
with ${\sfc(\eta,\ph)}$ defined again by \eqref{Sdef} and
\begin{equation}  
  s_k=\sqrt{\rho^2+\rho_\oix^2-2\rho\rho_\oix\cos\bigl(\Delta\ph+2\pi k/\iconpar\bigr)}
\end{equation}
being the distance from the line along the ${k}$-th geodesic orthogonal to the line.
Magnetic field ${\EMB=\EMBc^{\hat z}\tens{e}_{z}+\EMBc^{\hat \rho}\tens{e}_{\rho}+\EMBc^{\hat \ph}\tens{e}_{\ph}}$ 
is then given by
\begin{align}
\begin{split}
  \EMBc^{\hat\rho}&=-\frac{I}{2\pi}\sum_{k=\kin}^{\kfi}\frac{\rho_\oix\sin(\Delta\ph+2\pi k/\iconpar)}{s_k^2}\\
         &\mspace{-30mu}+\frac{\iconpar I}{8\pi^2}\int_0^{\infty}
         \frac1\rho\,\frac{\partial\sfc}{\partial\ph}(\eta,\Delta\ph)\,\ln\bigl(\rho^2+\rho_\oix^2+2\rho\rho_\oix\ch\eta\bigr)\,d\eta\commae
\end{split}\notag\\
\EMBc^{\hat z}&=0\commae\label{magfield}\\
\begin{split}
  \EMBc^{\hat\ph}&=\frac{I}{2\pi}\sum_{k=\kin}^{\kfi}\frac{\rho-\rho_\oix\cos(\Delta\ph+2\pi k/\iconpar)}{s_k^2}\\
         &\quad-\frac{\iconpar I}{4\pi^2}\int_0^{\infty}
         \frac{\sfc(\eta,\Delta\ph)\,(\rho+\rho_\oix\ch\eta)\,d\eta}{\rho^2+\rho_\oix^2+2\rho\rho_\oix\ch\eta}\period
\end{split}\notag
\end{align}
The magnetic lines for such a field are shown in Fig.~\ref{fig:maglines}. We observe
that they have the standard structure of the magnetic field around a straight
line current except that they are deformed near the string. We will discuss this deformation
more in the following section.

Similarly to the self-force \eqref{elselfforce} of the static charge  
the current line acts on itself by a self-force which is now, however, 
pushing the source towards the string. 
Subtracting the standard empty space contribution to 
the magnetic field at the position of the line 
the force on the line leads to
\begin{equation}\label{magselfforce}
  \tens{F}^{\,\slf}=-\frac{I^2}{4\pi\rho_\oix}\,(\iconpar-1)\;\tens{e}_\rho
\end{equation}
(here we used integral \eqref{intsfc} from the appendix).

\section{The field at large distances from the string}
\label{sc:farfield}

Expanding the potential \eqref{scpot} of 
a static charge at a large distance ${r\gg\rho_\oix}$
from the origin (located on the string, 
i.e., assuming ${\Delta z=r\cos\tht}$, ${\rho=r\sin\tht}$), 
we obtain
\begin{equation}\label{scpotfarexp}
\begin{split}
  \EMP=&\frac{\charge}{4\pi}\frac1r\sum_{k=\kin}^\kfi
  \Bigl[1+\frac{\rho_\oix}{r}\sin\tht\cos\Delta\ph_k
  +\OO\Bigl(\frac{\rho_\oix^2}{r^2}\Bigr)\Bigr]\\
  &-\frac{\iconpar\charge}{8\pi^2}\frac1r\int_0^\infty\!\!\!\!
  \sfc(\eta,\Delta\ph)\Bigl[1-\frac{\rho_\oix}{r}\sin\tht\ch\eta
  +\OO\Bigl(\frac{\rho_\oix^2}{r^2}\Bigl)\Bigr]d\eta\period
\end{split}\raisetag{9ex}
\end{equation}
Using integrals \eqref{intsfc} and \eqref{intchsfc} 
from Appendix~\ref{apx:intsfc} we get
\begin{equation}\label{scpotfarfield}
  \EMP=\frac1{4\pi}\frac{\iconpar\charge}r
  +\frac1r\,\OO\Bigl(\frac{\rho_\oix^2}{r^2}\Bigr)\period
\end{equation}
We see that the monopole contribution corresponds to 
the modified charge ${\iconpar\charge}$. The same result 
for the electric field ${\EME}$ can be obtained using the Gauss law: 
the density of the electric flux (the magnitude of ${\EME}$) 
at infinity has to be bigger to compensate 
the smaller area of a distant \vague{sphere} which,
due to the conicity of space, grows only as ${4\pi r^2/\iconpar}$.

An interesting feature of the far field is the absence 
of a dipole term---indeed, the term proportional 
to ${r^{-2}}$ is missing in \eqref{scpotfarfield}.
This behavior can be confirmed investigating 
the far field of the dipole located near the string. 
The expansion of the potential \eqref{eldip} for large ${r}$
leads to
\begin{equation}\label{eldipfarexp}
\begin{split}
  \EMP=&\frac1{4\pi}\frac1{r^2}\sum_{k=\kin}^{\kfi}\Bigl[
  \cos\tht\, p^{\hat z}+\sin\tht\cos\Delta\ph_k \, p^{\hat\rho}\\[-3ex]
  &\mspace{240mu}+\sin\tht\sin\Delta\ph_k\,p^{\hat\ph}\Bigr]\\
  &+\frac\iconpar{8\pi^2}\frac1r\int_0^\infty\frac1{\rho_\oix}\frac{\partial\sfc}{\partial\ph}(\eta,\Delta\ph)\,p^{\hat\ph}\,d\eta\\
  &+\frac\iconpar{8\pi^2}\frac1{r^2}\int_0^\infty\Bigl[\bigl(
  -\cos\tht\,p^{\hat z}+\sin\tht\ch\eta\,p^{\hat\rho}\bigr)\sfc(\eta,\Delta\ph)\\
  &\mspace{165mu}-\sin\tht\ch\eta\,\frac{\partial\sfc}{\partial\ph}(\eta,\Delta\ph)\,p^{\hat\ph}\Bigr]d\eta\\[-1ex]
  &+\frac1{r^2}\OO\Bigl(\frac{\rho_\oix}{r}\Bigr)\period
\end{split}\raisetag{20ex}
\end{equation}
Substituting for the integrals expressions \eqref{intsfc}, \eqref{intchsfc}, \eqref{intdsfc}, and \eqref{intchdsfc}
we obtain
\begin{equation}\label{eldipfarfield}
  \EMP=\frac1{4\pi}\frac{\iconpar\,p^{\hat z}\cos\tht}{r^2}
  +\frac1{r^2}\OO\Bigl(\frac{\rho_\oix}{r}\Bigr)\period
\end{equation}
We see that only the component ${p^{\hat z}}$ of the dipole parallel to the string contributes in the order ${r^{-2}}$.
The field of the dipole oriented \emph{orthogonally} to the string is suppressed at large distances; it
falls-off as ${r^{-3}}$.

It is surprising that the suppression of ${r^{-2}}$ term for the dipole orthogonal to the
string is non-smooth in the limit ${\iconpar\to1}$, i.e., in the limit of the Minkowski spacetime: 
the ${r^{-2}}$ term is present for ${\iconpar=1}$, however, it vanishes for
any ${\iconpar>1}$. It turns out that the Minkowski limit is achieved by an \vague{enlarging}
the field's near-by zone to infinity. 
In other words, a far zone, where the field is described by \eqref{scpotfarfield}
or \eqref{eldipfarfield}, starts at larger and larger distances with ${\iconpar}$ approaching ${1}$; 
and it disappears completely for ${\iconpar=1}$.

It is not very clear how to estimate  a \vague{size} of the far zone analytically;
however, a numerical analysis indicates that it \vague{shrinks} to infinity
faster than any power of a deficit angle parameter ${1/(\iconpar-1)}$. 
One way how to determine the size of the far zone is to transform it to a domain near 
the string, using a spherical inversion with a center on the string. 
Such a domain near the string will be studied in the next section---cf., 
e.g., Fig.~\ref{fig:fltdomsize} for an estimate of its size.

In the Minkowski spacetime the conformal transformation of the electromagnetic
field associated with the spherical inversion transforms 
the dipole field into a homogeneous field. The vanishing dipole term 
far from the string thus suggests a suppression  
of the homogeneous component of the field near the string. Let us
study this property in more detail.

\section{Suppression of the field of distant sources near the string}
\label{sc:homfield}

Expanding the scalar potential \eqref{scpot} of the static charge 
in the domain near the string, i.e., for ${\rho\ll\rho_\oix}$,
and using integrals \eqref{intdsfc}, \eqref{intchsfc} we obtain
\begin{equation}\label{scpotstrexp}  
\begin{split}
\EMP &= \frac\charge{4\pi}\frac1{\rho_\oix}\,\sum_{k=\kin}^{\kfi}
   \Bigl(1+\frac\rho{\rho_\oix}\cos\Delta\ph_k+\OO\Bigl({\frac{\rho^2}{\rho_\oix^2}}\Bigr)\Bigr)\\
  &\quad-\frac{\iconpar\charge}{8\pi^2}\frac1{\rho_\oix}\int_{0}^{\infty}\!\!
  \Bigl(1-\frac\rho{\rho_\oix}\,\ch\eta+
  \OO\Bigl({\frac{\rho^2}{\rho_\oix^2}}\Bigr)\Bigr)\,\sfc(\eta,\Delta\ph)\,d\eta\\
  &=\frac\charge{4\pi}\frac1{\rho_\oix}+\frac1{\rho_\oix}\,\OO\Bigl({\frac{\rho^2}{\rho_\oix^2}}\Bigr)\period
\end{split}\raisetag{4ex}
\end{equation}
Up to the linear order there is no term depending on a 
position---the scalar potential ${\EMP}$
is constant near the string and hence, 
the electric field ${\EME}$ is vanishing. 
This is a surprising result: it means that the field
of large distant charges is suppressed near the cosmic string.
In other words, it is possible to hide oneself near the string 
from the electric field of strong charges.

\begin{figure}
\includegraphics{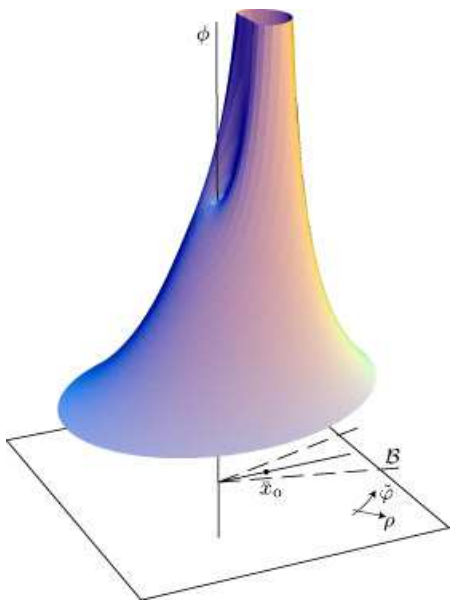}
\\
\caption{\label{fig:scpotflt}%
Scalar potential of the static charge near the string with
a \vague{large} deficit angle (namely, ${\iconpar=1.8}$).
We can se that it has a small plateau near the string. 
The electric field ${\EME}$ is suppressed here.
Such a plateau exists for any value ${\iconpar\neq1}$,
however, it is very small for ${\iconpar}$ close to ${1}$,
cf.\ Figs.~\ref{fig:scpot} and \ref{fig:fltdomsize}.}%
\end{figure}

\begin{figure}
\includegraphics{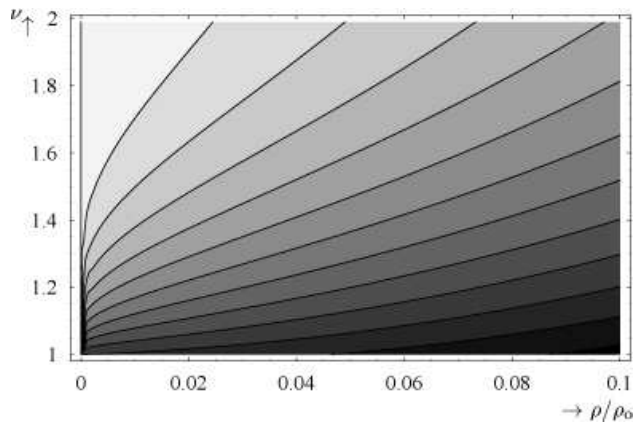}
\\
\caption{\label{fig:fltdomsize}%
The diagram illustrates the dependence of a typical 
size of the domain near the string, 
where the electric field ${\EME}$ of the static charge is suppressed,  
on the parameter ${\iconpar}$.
To estimate a size of the domain, 
a value of the field is evaluated along 
a line going through the charge orthogonally to the string. 
The horizontal direction in the figure corresponds
to the distance from the string.
The vertical direction corresponds to different values
of the parameter ${\iconpar}$.
The contours of constant values of the field
are drawn, with shading indicating the strength of the field
(the shading is drawn only in discrete steps given by contours).  
For ${\iconpar>1.5}$, the interval of small field (light shading)
is large; for ${\iconpar\to1}$ this intervals shrinks 
very rapidly and it disappears for ${\iconpar=1}$.
See also graphs of scalar potential for ${\iconpar=1.2}$ in Fig.~\ref{fig:scpot}(c)
and for ${\iconpar=1.8}$ in Fig.~\ref{fig:scpotflt}. Only in the latter one
a flat domain near the string can be easily identified.
A similar analysis applies also for magnetic field ${\EMB}$
discussed in Sec.~\ref{sc:magfield}. 
A domain of the expulsion of magnetic lines near the string 
can be identified, e.g., in Fig.~\ref{fig:maglines}.}%
\end{figure}

The expulsion of the field away from the string
is not continuous in the limit ${\iconpar\to1}$:
the electric field ${\EME}$ is negligible near the string
for any value ${\iconpar>1}$, but it is present for ${\iconpar=1}$.
The limit of no string is actually realized by 
a contraction of a domain around the string in which the field is suppressed.
For ${\iconpar}$ close to ${1}$ 
the domain of vanishing electric field is rather small.
It is nontrivial only for ${\iconpar\gtrsim3/2}$
(see Fig.~\ref{fig:scpotflt}), however it shrinks 
rapidly with ${\iconpar\to1}$, as it is seen in Fig.~\ref{fig:fltdomsize}.
A numerical study indicates that a typical size of this domain
decreases faster than any power of a deficit angle parameter ${\iconpar-1}$.

A similar discussion applies also for a magnetic field---e.g.,
the magnetic field discussed in Sec.~\ref{sc:magfield}
is also suppressed near the string.

The behavior of the field can be also rephrased 
as a nonexistence of a homogeneous electric or magnetic field 
perpendicular to the cosmic string.
Indeed, in empty Minkowski spacetime the homogeneous electric field 
can be constructed as a field of a large charge at large distances.
As we have seen, in the spacetime of cosmic string the field of
the charge located at large distances from the string is suppressed
and there is thus no analogue of the homogeneous field perpendicular to the string.
The same result can be obtained also by studying a general static field around the string 
(i.e., by a direct solution of the homogeneous Laplace equation
in the space with a deficit angle) \cite{Krtous:TPPKS}.

\section{Summary}

We have discussed the electromagnetic field in the spacetime of a cosmic string.
The retarded Green function \eqref{grfcexpl} for the vector potential
was found. It can be split into a \vague{geodetical} part \eqref{grfcgeod}
corresponding to the empty space propagation of the field, and into
a \vague{scattered} part \eqref{grfcscat} corresponding to scattering on the curvature
located within the string.

Using this Green function electric and magnetic fields of various 
simple static sources were constructed. It was demonstrated that although 
the spacetime is locally flat (outside the string), the global deficit angle 
causes several interesting phenomena. We found a self-interaction of the
point charge (repulsive away from the string), of the dipole (turning the dipole into 
the ${\tens{e}_{\ph}}$ direction \vague{around} the string), 
or of the line current (attractive toward the string).
In general, the self-interaction of charges and currents near the string
implies that the string participates on electromagnetic processes even
if it is not charged itself. It can have consequences when studying 
the motion of cosmic strings through, e.g., ionised plasma in an early Universe.

Further, we found that the field of a static source at 
large distances has a different multipole structure,
namely, there is no dipole term corresponding 
to the dipole perpendicular to the string.

Near the string, the electric and magnetic fields orthogonal to the string are suppressed.
For the same reason, there exists no homogeneous static field perpendicular to the string.

Most of the discussed effects can easily be understood in the special
case ${\iconpar=2}$ (respectively, for ${\iconpar}$ being any integer). 
In this case the scattered part of the field vanishes and the resulting
field can be obtained as the field in the half of Minkowski spacetime
given by the real source and by a fictitious source obtained
as a reflection of the real source with respect to the axis
(these two contributions correspond to different 
terms in the sum over geodetics in \eqref{scpot}).
The self-interaction is thus given by the interaction of the
real and the fictitious sources. A suppression of the dipole field
at large distances arises because a combination of 
the dipole (orthogonal to the string) and of its 
reflected image gives a quadrupole configuration. 
And finally, the expulsion of the field away from the string
arises as a cancelation of the fields of the source and of its
image in middle between them.

After deriving the general form of the Green function we concentrated 
mainly on static phenomena.
As a next step it would be interesting to study similar questions in dynamical contexts;
for example, the scattering of plane waves on the string or the field
of a strong oscillating dipole located far from the string.

\vspace{-1.5ex}
\begin{acknowledgments}
The author would like to thank Prof. Gal'tsov for a guidance during author's study stay
at Lomonosov University in Moscow back in 1988, during which most of the 
work \cite{Krtous:TPPKS} was done, and 
Prof. Bi\v{c}\'ak for arranging this stay and for reading both
\cite{Krtous:TPPKS} and the manuscript of this paper.
\end{acknowledgments}
\vspace{-1.5ex}

\appendix

\section{Integration of the Green function}
\label{apx:grfc}

In this Appendix we evaluate the integral \eqref{grfcint}. 
The choice of the integration contour in
the complex plane of ${\omega}$ guarantees that the integral
is vanishing for ${\Delta t<0}$. For ${\Delta t>0}$ we can close the contour 
in the upper half of the complex plane and the integration over ${\omega}$ leads to residues 
at simple poles ${\omega=\pm\sqrt{\kappa^2+\lambda^2}}$. This gives
\begin{equation}  
\begin{split}
&\grfc(x|x')= -\frac{i\iconpar}{(2\pi)^2}\;\theta(\Delta t)
\sum_{n\in\integern}e^{-i n \iconpar \Delta\ph}\\
&\quad\times\int_0^\infty \!\!\lambda d\lambda\; \BesJ{\abs{\iconpar n+\sg}}(\lambda\rho)\,\BesJ{\abs{\iconpar n+\sg}}(\lambda\rho')
\int_{-\infty}^\infty\frac{d\kappa}{\sqrt{\lambda^2+\kappa^2}}\\
&\qquad\quad\times\frac12\Bigl(e^{i\lambda\bigl(\sqrt{1+\frac{\kappa^2}{\lambda^2}}\Delta t-\frac\kappa\lambda\Delta z\bigr)}
-e^{-i\lambda\bigl(\sqrt{1+\frac{\kappa^2}{\lambda^2}}\Delta t+\frac\kappa\lambda\Delta z\bigr)}\Bigr)\period
\end{split}\raisetag{16ex}
\end{equation}
Next we substitute for ${\kappa}$
\begin{equation}  \label{alphasubst}
\sh\alpha=\frac\kappa\lambda\comma\ch\alpha=\sqrt{1+\frac{\kappa^2}{\lambda^2}}\period
\end{equation}
For ${\Delta z^2>\Delta t^2}$ we introduce the notation
\begin{equation}
\Delta t = s \sh\beta\comma \Delta z= s \ch\beta\comma \sign s =\sign\Delta z\commae
\end{equation}
and we obtain
\begin{equation}  
\begin{split}
&\grfc(x|x')= \frac{\iconpar}{(2\pi)^2}\;\theta(\Delta t)\,\theta\bigl(\Delta z^2\!\!-\!\Delta t^2\bigr)
\sum_{n\in\integern}e^{-i n \iconpar \Delta\ph}\\
&\qquad\quad\times\int_0^\infty \!\!\lambda d\lambda\; \BesJ{\abs{\iconpar n+\sg}}(\lambda\rho)\,\BesJ{\abs{\iconpar n+\sg}}(\lambda\rho')\\
&\qquad\qquad\times\int_{-\infty}^\infty \!\!d\alpha\;\frac1{2i}\,\Bigl(e^{-i\lambda s\sh(\alpha-\beta)}-e^{-i\lambda s\sh(\alpha+\beta)}\Bigr)\period
\end{split}\raisetag{12ex}
\end{equation}\pagebreak[2]
By a simple shift of ${\alpha}$ we get
\begin{equation}  
\grfc(x|x')= 0\qquad\text{for}\quad \Delta z^2>\Delta t^2\period
\end{equation}
For  ${\Delta t^2>\Delta z^2}$, ${\Delta t>0}$ the substitution \eqref{alphasubst} and
\begin{equation}
\Delta t = s \ch\beta\comma \Delta z= s \sh\beta\commae
\end{equation}
lead to
\begin{equation}\label{grfcscatapx}
\begin{split}
&\grfc(x|x')= \frac{\iconpar}{(2\pi)^2}\;\theta\bigr(\Delta t\bigl)\theta\bigl(\Delta t^2-\Delta z^2\bigr)
\sum_{n\in\integern}e^{-i n \iconpar \Delta\ph}\\
&\quad\times\int_0^\infty \!\!\!\!\lambda d\lambda\; \BesJ{\abs{\iconpar n+\sg}}(\lambda\rho)\,\BesJ{\abs{\iconpar n+\sg}}(\lambda\rho')
\int_{-\infty}^\infty \!\!\!\!d\alpha\,\sin\bigl(\lambda s\ch\alpha\bigr)\period
\end{split}\raisetag{9ex}
\end{equation}
The integration over ${\alpha}$ gives the Bessel function ${\pi\BesJ{0}(\lambda s)}$
(see 3.996.4 of \cite{GradshteinRyzhik:book}). We are thus left with integration over three 
Bessel functions.

With help of 6.578.8 and 8.754 of \cite{GradshteinRyzhik:book} (cf.\ also \cite{Macdonald:1909})
we see that for ${\rho,\rho',s>0}$, and ${\re\eps>-1}$ such integration gives
\begin{equation}  \label{JJJint}
\begin{split}
  &\int_0^\infty\!\!\lambda d\lambda\,\BesJ{0}(s\lambda)\,\BesJ\eps(\rho\lambda)\,\BesJ\eps(\rho'\lambda)\\
  &\;=
  \begin{cases}
  &\!\!{\displaystyle-\frac{\sin(\eps\pi)\,\exp(-\eps\eta)}{\pi\,\rho\rho'\sh\eta}}\comma
     \text{where}\;\ch\eta=\frac{s^2-\rho^2-\rho'^2}{2\rho\rho'}\\
  &\mspace{160mu}\text{for}\;\rho+\rho'<s\commae\\[2ex]
  &\!\!{\displaystyle\frac{\cos(\eps\eta)}{\pi\,\rho\rho'\sin\eta}}\comma
     \text{where}\;\cos\eta\!=\!\frac{\rho^2+\rho'^2-s^2}{2\rho\rho'}\;,\;\eta\!\in\!(0,\pi)\\
  &\mspace{160mu}\text{for}\;\abs{\rho-\rho'}<s<\rho+\rho'\commae\\[2ex]
  &\!\!0\mspace{139mu}\quad\text{for}\;s<\abs{\rho-\rho'}\period
  \end{cases}
\end{split}\raisetag{27.5ex}
\end{equation}

We start with the case ${\rho+\rho'<s=\sqrt{\Delta t^2-\Delta z^2}}$. The condition
${\rho+\rho'=\sqrt{\Delta t^2-\Delta z^2}}$ means that the point ${x}$ can be connected 
with ${x'}$ by a null geodesic going from ${x'}$ to ${y}$ at the string and then
again by a null geodesic from ${y}$ to ${x}$ (cf.\ Fig.~\ref{fig:scat}). Inequality 
then allows that the points ${y}$ and ${x}$ are connected by a timelike geodesic.
The point ${x}$ belongs thus to the causal future of the point ${y}$ on the string
where the electromagnetic field propagating from ${x'}$ by the speed of light is scattered.
Using \eqref{JJJint} the Green function takes form
\begin{equation}  
\begin{split}
\grfcpart{\scat}(x|x')= &-\frac{\iconpar}{4\pi^2}\;
\frac{\theta\bigr(\Delta t\bigl)\theta\bigl(\Delta t^2\!-\!\Delta z^2\!-\!(\rho\!+\!\rho')^2\bigr)}{\rho\rho'\sh\eta}\\
&\times\sum_{n\in\integern}e^{-i n \iconpar \Delta\ph}\sin\bigl(\abs{\iconpar n +\sg}\pi\bigr) e^{-\abs{\iconpar n +\sg}\eta}\commae
\end{split}\raisetag{8.5ex}
\end{equation}\pagebreak[2]
with ${\eta}$ given by \eqref{etadef}. Summing geometrical series and 
doing some algebraic manipulations leads to 
the scattered part of the Green function:
\begin{equation}  
\begin{split}
  &\grfcpart{\scat}(x|x') = 
    -(-1)^\sg\frac{\iconpar}{8\pi^2}\frac{\theta(\Delta t)\,\theta\bigl(\Delta t^2\!-\!\Delta z^2\!-\!(\rho\!+\!\rho')^2\bigr)}{\rho\rho'}\\
    &\times\!\!\biggl[\frac{\ch\sg\eta}{\sh\eta}\Bigl(
    \frac{\sin\iconpar(\pi-\Delta\ph)}{\ch\iconpar\eta\!-\!\cos\iconpar(\pi\!-\!\Delta\ph)}
    \!+\!\frac{\sin\iconpar(\pi+\Delta\ph)}{\ch\iconpar\eta\!-\!\cos\iconpar(\pi\!+\!\Delta\ph)}\Bigr)\\
    &\;\;+i\sg\Bigl(\frac{\sh\iconpar\eta}{\ch\iconpar\eta\!-\!\cos\iconpar(\pi\!-\!\Delta\ph)}
    -\frac{\sh\iconpar\eta}{\ch\iconpar\eta\!-\!\cos\iconpar(\pi\!+\!\Delta\ph)}\Bigr)\biggr]\!\period
\end{split}\raisetag{0ex}
\end{equation}\\[2ex]

Applying \eqref{JJJint} for ${\abs{\rho-\rho'}<s<\rho+\rho'}$  we get 
(omitting for a moment step functions corresponding for these conditions)
\begin{equation}  
\grfcpart{\geod}(x|x')= \frac{\iconpar}{4\pi^2}\;
\frac{\theta\bigr(\Delta t\bigl)}{\rho\rho'\sin\eta}
\sum_{n\in\integern}e^{-i n \iconpar \Delta\ph}\cos\bigl((\iconpar n +\sg)\eta\bigr) 
\end{equation}
with 
\begin{equation}  
\cos\eta\!=\!\frac{-\Delta t^2+\Delta z^2+\rho^2+\rho'^2}{2\rho\rho'}\comma 
\eta\!\in\!(0,\pi)\period
\end{equation}
Summing Fourier series leads to the sum of delta-functions. Finally, if we change arguments of these
delta-functions, we find
\begin{widetext}
\begin{equation}\label{grfcgeodapx}
\begin{split}
\grfcpart{\geod}(x|x')&= \frac{\iconpar}{4\pi}\;
\frac{\theta\bigr(\Delta t\bigl)}{\rho\rho'\sin\eta}
\sum_{k\in\integern}\Bigl[e^{i\sg\eta}\,\delta\Bigl(\iconpar(\Delta\ph\!-\!\eta)\!+\!2\pi k\Bigr) 
+e^{\!-i\sg\eta}\,\delta\Bigl(\iconpar(\Delta\ph\!+\!\eta)\!+\!2\pi k\Bigr)\Bigr]=\\
&=\frac{\theta\bigr(\Delta t\bigl)}{2\pi}\mspace{-10mu}
\sum_{\substack{k\;\text{such that}\\\Delta\ph+2\pi k/\iconpar\in(-\pi,\pi)}}\mspace{-20mu}
e^{i\sg(\Delta\ph+\frac{2\pi k}{\iconpar})}\;\delta\Bigl(-\Delta t^2+\Delta z^2+\rho^2+\rho'^2
-2\rho\rho'\cos\bigl(\Delta\ph+2\pi k/\iconpar\bigr)\Bigr)\period
\end{split}
\end{equation}
\end{widetext}
Checking the support of the delta-functions,  we justify the omission of the step functions enforcing conditions
${\abs{\rho-\rho'}<s<\rho+\rho'}$.
Introducing notation \eqref{rkdef} we obtain the geodesic part of the Green function.

Adding \eqref{grfcgeodapx} and \eqref{grfcscatapx} we prove that the Green function ${\grfc(x|x')}$
has the form \eqref{grfcexpl}.

\section{Integrals of function ${S(\eta,\ph)}$}
\label{apx:intsfc}

In the main text we need to evaluate integrals ${\int\sfc\, d\eta}$, ${\int \ch\eta\,\sfc\, d\eta}$,
and ${\int \partial\sfc/\partial\ph\, d\eta}$, ${\int \ch\eta\, \partial\sfc/\partial\sfc\, d\eta}$. 
We first note that conditions \eqref{kinfi} are equivalent to the conditions
\begin{equation}  
\begin{aligned}
  \alpha_\iix &= \iconpar(\pi+\Delta\ph)+2\pi \kin-\pi\in(-\pi,\pi)\commae\\
  \alpha_\fix &= \iconpar(\pi-\Delta\ph)-2\pi \kfi-\pi\in(-\pi,\pi)\commae
\end{aligned}
\end{equation}
where, similarly to ${\kin}$ and ${\kfi}$, we do not write the dependence 
of ${\alpha_\iix}$ and ${\alpha_\fix}$ on ${\Delta\ph}$ explicitly.
With such defined ${\alpha_\iix}$ and ${\alpha_\fix}$ we have
\begin{equation}  
\sfc(\eta,\Delta\ph) 
  = -\frac{\sin\alpha_\iix}{\ch\iconpar\eta+\cos\alpha_\iix}
  - \frac{\sin\alpha_\fix}{\ch\iconpar\eta+\cos\alpha_\fix}\period
\end{equation}

Now we can use the result 3.514.1 from \cite{GradshteinRyzhik:book},
to find
\begin{equation}\label{intsfc}  
\int_0^\infty\sfc(\eta,\Delta\ph)\,d\eta=\frac{2\pi}{\iconpar}\bigl(-\iconpar+\kfi-\kin+1\bigr)\period
\end{equation}
Similarly, with help of 3.514.2 from \cite{GradshteinRyzhik:book} we get
\begin{equation}  
\begin{split}
&\int_0^\infty\ch\eta\,\sfc(\eta,\Delta\ph)\,d\eta=
   -\frac{\pi}{\iconpar\sin(\pi/\iconpar)}\\
   &\quad\times\!\Bigl(\sin\bigl(\Delta\ph\!+\!{\textstyle\frac{2\pi \kfi\!+\!\pi}{\iconpar}}\bigr)-
   \sin\bigl(\Delta\ph\!+\!{\textstyle\frac{2\pi \kin\!-\!\pi}{\iconpar}}\bigr)\Bigr)\period
\end{split}
\end{equation}
This can be rewritten as
\begin{equation}  \label{intchsfc}
\int_0^\infty\ch\eta\,\sfc(\eta,\Delta\ph)\,d\eta=
   -\frac{2\pi}{\iconpar}
   \sum_{k=\kin}^{\kfi} \cos\bigl(\Delta\ph+2\pi k/\iconpar\bigr)\period
\end{equation}

The derivative of ${\sfc(\eta,\Delta\ph)}$ with respect to the second argument is
\begin{equation}\label{dsfc}
\frac{\partial\sfc}{\partial\ph}\bigl(\eta,\Delta\ph\bigr)
  = -\frac{1+\ch\eta\cos\alpha_\fix}{(\ch\iconpar+\cos\alpha_\fix)^2}
    +\frac{1+\ch\eta\cos\alpha_\iix}{(\ch\iconpar+\cos\alpha_\iix)^2}\period
\end{equation}
As limiting cases of the integral 3.514.3 from \cite{GradshteinRyzhik:book} we get
\begin{equation}  \label{int3}
\begin{gathered}
\int_0^\infty\frac1{(\ch\iconpar\eta+\cos\alpha)^2}\,d\eta=\frac1\iconpar\frac{\sin\alpha-\alpha\cos\alpha}{\sin^3\!\alpha}\commae\\
\int_0^\infty\frac{\ch\eta}{(\ch\iconpar\eta+\cos\alpha)^2}\,d\eta=\frac1\iconpar\frac{\alpha-\sin\alpha\cos\alpha}{\sin^3\!\alpha}\commae
\end{gathered}
\end{equation}
with ${\alpha\in(0,\pi)}$. Combining these integrals we obtain
\begin{equation}
\int_0^\infty\frac{1+\ch\eta\cos\alpha}{(\ch\iconpar\eta+\cos\alpha)^2}\,d\eta=\frac1\iconpar\commae
\end{equation}
which holds for arbitrary ${\alpha}$.
It immediately follows that
\begin{equation}  \label{intdsfc}
\int_0^\infty\frac{\partial\sfc}{\partial\ph}(\eta,\Delta\ph)\,d\eta=0\period
\end{equation}

Rewriting ${\ch\eta\ch\iconpar\eta}$ as ${\ch(\iconpar-1)\eta+\sh\eta\sh\iconpar\eta}$
we can use the integrals 3.514.3 and 3.514.4 of \cite{GradshteinRyzhik:book}
to derive
\begin{equation} \label{inttmp} 
\int_0^\infty\!\ch\eta\,\frac{1+\ch\iconpar\eta\,\cos\alpha}{(\ch\iconpar\eta+\cos\alpha)^2}\,d\eta
=\frac{\pi}{\iconpar^2}\,\cos\frac\alpha\iconpar\commae
\end{equation}
with ${\alpha\in(0,\pi)}$.
Finally, Eqs.~\eqref{dsfc} and \eqref{inttmp} leads to
\begin{equation}
\begin{split}
&\int_0^\infty\ch\eta\,\frac{\partial\sfc}{\partial\ph}\bigl(\eta,\Delta\ph\bigr)\,d\eta
=\frac\pi{\iconpar\sin\frac\pi\iconpar}\\
&\quad\times\Bigl(\cos\Bigl(\Delta\ph\!+\!\frac{2\pi \kfi\!+\!\pi}\iconpar\Bigr)
- \cos\Bigl(-\Delta\ph\!+\!\frac{-2\pi \kin\!+\!\pi}\iconpar\Bigr)\Bigr)\commae
\end{split}\raisetag{9ex}
\end{equation}
which can be rewritten as
\begin{equation}\label{intchdsfc}
\int_0^\infty\ch\eta\,\frac{\partial\sfc}{\partial\ph}\bigl(\eta,\Delta\ph\bigr)\,d\eta
=\frac{2\pi}{\iconpar}\sum_{k=\kin}^{\kfi} \sin\Bigl(\Delta\ph+\frac{2\pi \kfi}\iconpar\Bigr)\period
\end{equation}


\end{document}